# Compact fully monolithic optomechanical accelerometer

Felipe Guzman[1,2,*], Lee M. Kumanchik[1], Ruven Spannagel[1], Claus Braxmaier[1]

[1]*German Aerospace Center (DLR) – Institute of Space Systems & University of Bremen, Linzer Str. 1, 28359 Bremen, Germany*
[2]*College of Optical Sciences, University of Arizona, Tucson, AZ 85721, USA*
*\* felipe@optics.arizona.edu*

**Abstract:** We present a highly compact and fully monolithic optomechanical accelerometer fabricated of a single wafer of fused-silica with a total volume of less than 2 cm$^3$ and a total mass of approximately 4 grams. This sensor demonstrates an unprecedented *mQ*-product of 50 kg, highlighting its unique capabilities for reaching high sensitivity in acceleration sensing. The sensor includes a monolithically co-fabricated Fabry-Pérot interferometer that uses the accelerometer test mass as a mirror to measure its displacement in orders of 10$^{-12}$ m/√Hz, reaching acceleration sensitivities to the order of 100 ng/√Hz – 1 µg/√Hz above 1 Hz.

## 1. Introduction

A wide variety of scientific observations and industrial applications rely on the performance of dynamic measurement instruments, such as accelerometers. Particularly, applications in precision measurements require, at their core, sensors capable of measuring with extremely high sensitivity and often with wide bandwidth, either vibrations and acoustics, seismic or spurious forces acting on their test platform. A new window into micro-mechanical sensors and micro-optical sensing has opened in recent years with modern optomechanics, where the combination of low loss and highly stable, carefully designed and manufactured micro-mechanics are combined with micro-optical sensors of outstanding sensitivity, enabling us to achieve unprecedented performances in comparison to conventional technologies.

Making use of established MEMS fabrication technologies various concepts have been developed recently [1–8], some of which include advanced photonic and micro-optical systems with interesting detection techniques. The acceleration sensitivities reported range typically near or above 1 µ*g*/√Hz at relatively high frequencies in the few hundred Hz to several kHz.

Building upon previous research work on optomechanical acceleration and force sensors [9–12], we present the development of a highly compact and fully monolithic optomechanical acceleration sensor capable of reaching µ*g*/√Hz sensitivities down to 100 mHz. This sensor is made of fused-silica, which is a low loss material at room temperature and has a very low coefficient of thermal expansion (CTE). The sensor has a footprint of approximately 2×2×0.5 cm with a total mass of approximately 4 grams, and reaches an unprecedented *mQ*-product of 50 kg, which is essential for high sensitivity acceleration sensing. Previous developments had demonstrated *mQ*-products in the range of 1–2 kg [9].

In addition to the improvements in acceleration metrology and calibrations, this research will greatly impact enabling technologies, particularly for inertial navigation systems. Current performance challenges, such as bias, scale factor stability, and noise floor, are tied to the uncertainty of the inertial navigation systems, which are aspects that greatly benefit from fully monolithic sensors made of highly stable materials. Similarly, research in seismology and ground- and space-based geodesy will benefit from highly compact devices of significantly higher performance.

## 2. Optomechanical accelerometer

Acceleration resolution is fundamentally limited by thermal fluctuations of the test mass, which are given for a simple harmonic oscillator (SHO) at high temperature by Equation 1, where *m* is the test mass, *Q* is the mechanical quality factor, *T* is the temperature of the test mass thermal bath, $k_B$ is the Boltzmann constant, and $\omega_0$ is the natural frequency of the mechanical oscillator. The stiffness of the mechanical oscillator has been intentionally designed to be high for very small test mass deflection amplitudes, *δx*, maintaining displacement well within the linear regime where the SHO approximation holds extremely well.

$$\underbrace{\delta a}_{SHO} = \sqrt{\frac{4 k_B T \omega_0}{m\, Q}} \quad (1), \qquad \frac{x(\omega)}{a(\omega)} = -\frac{1}{\omega_0^2 - \omega^2 + \frac{i \omega_0}{Q}\omega} \quad (2),$$

$$\delta a = \omega_0^2 \underbrace{\delta x}_{\substack{\text{displacement}\\ \text{noise density}}} \quad (3)$$

Typically, acceleration is obtained from a direct displacement measurement, and both observables, *x*(*ω*) and *a*(*ω*), are related by the transfer function in Equation 2. At the low frequency (DC) limit, significantly below $\omega_0$, acceleration and displacement follow the relation outlined by Equation 3.

In this work, we achieved a sensor assembly that is significantly more compact, with a much lower total mass and with much better performance through the fabrication of a fully monolithic sensor that comprises the supporting base, a flexing mass of 109 milligrams, and optical measurement Fabry-Pérot interferometer (FPI) mirrors from a single wafer. This sensor exhibits a very high mechanical Q of 4.6×105 and a *mQ*-product of 50 kg, which is a key figure of merit for very low loss acceleration sensitivity. It is noteworthy that the high *Q* is achieved in spite of how the supporting base is mounted to the superstructure (mechanical ground); in this case by a simple post clamp with a nylon-tipped set screw. So-called mounting loss depends on the details of the flexure-to-ground path and is typically lumped into the leftover loss after other losses are calculated [13]. The wide surface of the supporting base helps redistribute stress over the mounting interface, in turn reducing vulnerability to the nature of the clamp that is very important in transferring the performance reported here to real-world installations.

Figure 1 shows a CAD image of the fully monolithic optomechanical accelerometer, including a simple modal analysis of its first 3 resonances.

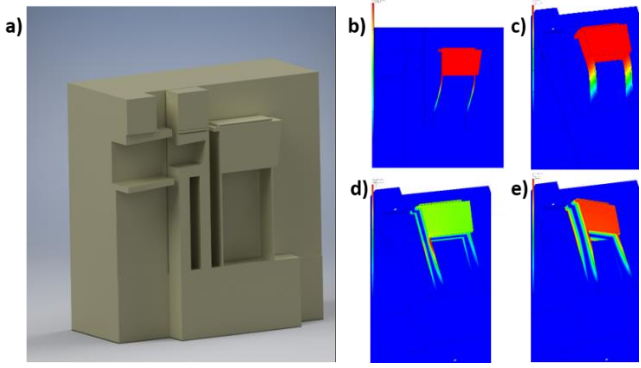

*Figure 1. a) CAD model of a fully monolithic optomechanical accelerometer. b) and c) Modal analysis and illustration of test mass dynamics for the main mechanical resonance $\omega_o$=689.5 Hz. d) and e) Modal analysis and illustration of test mass dynamics for mechanical modes $\omega_1$=3.85$\omega_o$ and $\omega_2$=12.8$\omega_o$, respectively.*

One of the advantages of this geometry is the ability to constrain higher order modes to frequencies significantly higher than the main resonance frequency, which defines the observation bandwidth. The stiffness of the mechanical oscillator is directly proportional to $\omega_o^2$ and the effective mass is the same for the following first mode, $\omega_1$. According to the modal analysis, the frequency of this first mode, $\omega_1$=3.85$\omega_o$, is nearly four times higher than the fundamental mode, $\omega_o$, which means that the amplitude of test mass deflections at $\omega_1$ are over an order of magnitude smaller for acceleration components orthogonal to the main sensitive axis. The color of the figures b) – e), red to blue, indicate the amplitude of the test mass displacement along the FPI axis, where red shows larger amplitudes. This shows a further reduction in susceptibility to orthogonal displacements, particularly for a flat-flat cavity DC readout, since these displacements do not couple strongly to the measurement. While all three modes shown in Fig. 1 are related, each of them can be tuned by different parameters: a) the fundamental mode, $\omega_o$, is strongly defined by the thickness of the leaf springs, b) the height of the leaf springs defines the first mode, $\omega_1$, and c) the second mode, $\omega_2$, is mainly characterized by the spacing between the two leaf springs.

To build an integrated high sensitivity optical sensor, we designed a plano-plano FPI within the fused-silica structure that is jointly fabricated with the mechanical oscillator from a single wafer. A flat surface within the rigid structure is separated by approximately 1 mm from the monolithically fabricated oscillating test mass. This flat side of the test mass and the flat surface build a FPI that is used as the optical displacement sensor.

Previous research on similar devices [7–12,14–15] demonstrated fiber-based and free-beam FPIs at various finesse levels, from 2 to 10^5. In this work, due to the fully monolithic fabrication of this sensor, we decided to operate the FPI at a finesse of 2 and a nominal FPI length of 1 mm, resulting in a free spectral range (FSR) of approximately 150 GHz with a broad linewidth of approximately 75 GHz. The FPI is read out with a DC scheme, where the test mass displacement $\delta x$, which is equivalent to FPI length changes, induces corresponding intensity changes in the reflected optical field. The optical FPI output measured at a photodetector $V(L,\lambda)$ can be expressed by an Airy function in terms of the laser wavelength $\lambda$ as

$$V(L,\lambda) = v_o + \gamma \frac{(1+R)^2}{2}\left[\frac{1-\cos\left(\frac{4\pi}{\lambda}L\right)}{1+R^2+2R\cos\left(\frac{4\pi}{\lambda}L\right)}\right] \quad (4),$$

$$\delta x = \frac{L}{\lambda_o}\left(\frac{\partial V(L,\lambda_o)}{\partial \lambda}\right)^{-1}\delta V \quad (5).$$

where *L* is the FPI length, $\lambda$ is the laser wavelength, *R* is the net reflectivity of the FPI mirrors, $\gamma$ is the signal visibility, and $v_o$ is an offset. When the laser wavelength is tuned to the maximum sensitivity points of the FPI, $\lambda_o$, the test mass displacement, $\delta x$, that corresponds to length fluctuations of the FPI can be observed through fluctuations of the output voltage, $\delta V$, measured at the photodetector, as given in Equation 5.

The fully monolithic topology allows us to inject a laser beam into the FPI either as a free detached beam, or, for convenience, through a GRIN collimator that can be mounted on the sensor itself. A generic pocket has been foreseen in the sensor design for the latter and, given the multiple dimensions available in the market for these types of collimators, it is also possible to incorporate a custom-built adapter that mates the collimator to the pocket. Since the FPI mirrors are a monolithic part of the sensor itself, either attachment of a GRIN collimator or the injection of an external beam into the FPI will not affect the length sensitivity or stability of the optical measurement in first order. One possible secondary effect would be severe coupling jitter of the input beam into the FP, impacting the performance of the optical measurement. However, the coupling stability is such that this effect has not been observed.

In this paper we present the results measured with the monolithic sensor when using a fiber GRIN collimator for convenience. However, we successfully tested the injection of a free laser beam from a detached source into the monolithic flexible FPI as well, showing similar results.

Figure 2 shows photographs of the sensor assembly itself, including the fiber GRIN collimator and its adapter.

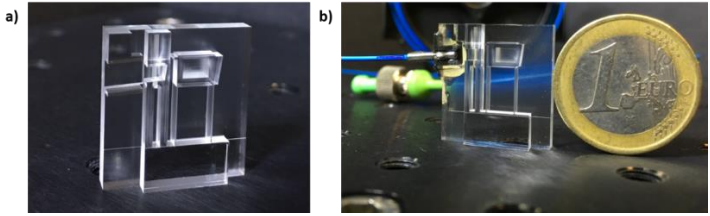

*Figure 2. a) Photograph of a fully monolithic optomechanical accelerometer. b) Photograph of sensor with an incorporated GRIN collimator and 3D-printed adapter next to a 1 Euro coin.*

Figure 3 outlines the experimental setup, optic and fiber-optic circuitry used in these investigations.

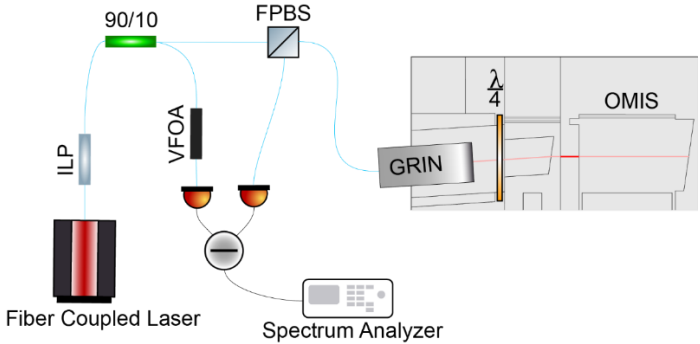

*Figure 3. Diagram of experimental apparatus to characterize and operate the monolithic optomechanical accelerometer. We use a tunable external cavity laser that is fiber-coupled and connected to a fiber in-line polarizer (ILP). The output of the ILP goes into a 90/10 coupler, which sends 90% of the light via a fiber polarizing beam splitter, a fiber-coupled GRIN collimator and a quarter-wave-plate into the monolithic FPI. The remaining 10% go through a variable fiber optical attenuator (VFOA) onto a photodetector. The reflected optical field from the FPI is coupled back into the GRIN collimator and exits the FPBS through a separate output that is connected to a photodetector. The output of the two photodetectors is subtracted to cancel out common-mode noise, and recorded by a spectrum analyzer.*

We use a largely tunable laser (90 nm tunable range) centered around 1565 nm to measure and characterize the optical response of the FPI over several FSRs, thus having a direct measurement of the optical transduction to test mass displacement, $\frac{\partial V(L,\lambda_o)}{\partial \lambda}$, as given in Equation 5. The large tunable range of the laser is also advantageous to tune the laser to high sensitivity wavelength points of the FPI, $\lambda_o$, which can be identified through a direct FSR measurement.

The laser beam is directly coupled into a fiber optic circuit. This circuit consists of an in-line polarizer (ILP) connected to a 90/10 coupler: a) 10% of the light is transmitted through a variable fiber optical attenuator (VFOA) and a photoreceiver that is used for purposes of intensity noise cancellation; b) the second branch directs the remaining 90% of the light through a fiber polarizing beam splitter (FPBS) and injects the light into the FPI through a GRIN collimator and a quarter-wave-plate. Angled surfaces near the input and output of the FPI scatter away unwanted Fresnel reflections. The reflected beam coming from the FPI is coupled back into the fiber through the GRIN lens and is directed by the FPBS to a photodetector for the measurement.

The entire fiber-optic apparatus and optomechanical sensor are mounted on a passive vibration isolation platform (corner frequency ~10 Hz) inside a vacuum chamber that runs nominally at a pressure of approximately $10^{-4}$ mBar. This vacuum chamber is installed on top of an optical table that is pneumatically suspended for a second level of vibration isolation (corner frequency ~4 Hz).

## 3. Mechanical and optical characteristics

The characteristics of the compact mechanical oscillator are obtained through a ringdown measurement by injecting an external excitation through a piezo-electric transducer (PZT) and recording the oscillator decay response upon switching off this excitation. At this point, the oscillator rings down at its natural frequency, $\omega_o=2\pi f_o$, where $f_o$=689.5 Hz, which is easily determined by spectral analysis of the recorded time series. The mechanical quality factor, $Q$=463.24k, is obtained from a fit to the exponential decay corresponding to the resulting envelope of the recorded time series upon demodulation at the frequency $\omega_o$. Figure 4 shows the time series and results obtained from this ringdown measurement.

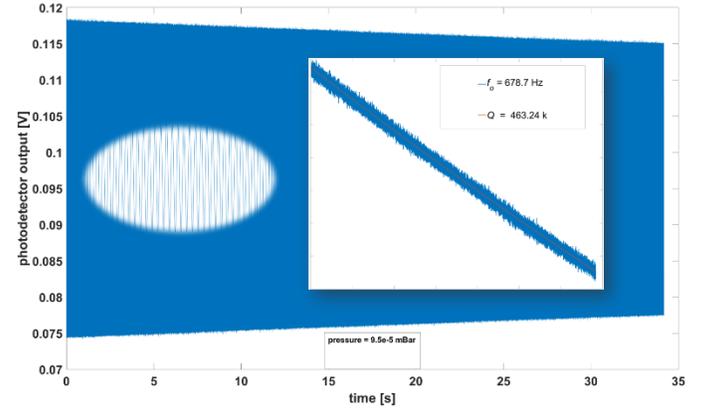

*Figure 4. Ringdown measurement of mechanical oscillator measured through the co-fabricated optical Fabry-Pérot. A resonance frequency $\omega_o$=2$\pi$ 689.5 rad/s was measured with a mechanical quality factor, Q=463.24k.*

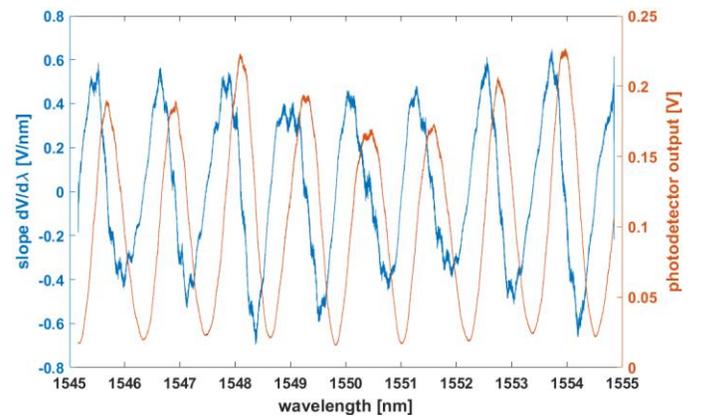

*Figure 5. Wavelength scan of co-fabricated optical Fabry-Pérot sensor. The left vertical axis shows the sensor's sensitivity/slope to FPI length driven wavelength/frequency changes. The right vertical axis shows the FPI reflection spectrum – free spectral range – from 1545 nm to 1555 nm. From this plot we determine a free spectral range of approximately 146 GHz, which corresponds to a FPI length of approximately 1 mm.*

The monolithic FPI sensor has been designed to have a length of approximately 1 mm. By using a largely tunable laser we have measured the reflection spectrum of this sensor and can also directly determine the sensor's sensitivity (slope) to displacement. Figure 5 shows the results from these FPI spectroscopy measurements.

The sensitivity trace (in blue) from Figure 5 displays a transduction factor of approximately $\frac{\partial V(L,\lambda_o)}{\partial \lambda} = 5\times10^8$ V/m, when the laser wavelength is tuned to the wavelength points of high sensitivity in the monolithic FPI.

## 4. Results

The sensor was placed in a vacuum chamber operated at a pressure below $10^{-4}$ mBar. The laser light was injected as a free beam into the monolithic FPI using the GRIN collimator connected to the fiber circuitry shown in Fig. 3. Due to the low reflectivity of the monolithic mirrors, most of the optical power is lost, however, the FPI operates with approximately 800 µW of optical power, which is reflected back into the fiber system for detection.

Figure 6 shows a linear spectral density of the test mass displacement. At around 1 Hz we obtain a sensitivity of 2 pm/√Hz with a decreasing trend down to 0.1 pm/√Hz at 1 kHz. Our test mass displacement measurement reaches a level of approximately 10 pm/√Hz slightly below 100 mHz.

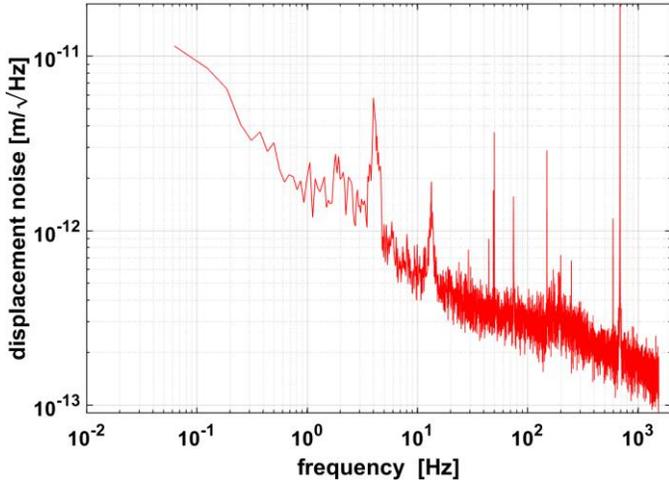

Figure 6. Test mass displacement noise of the optomechanical accelerometer measured with the co-fabricated monolithic FPI.

Ancillary measurements show that this excess noise is mostly dominated by environmentally induced noise in the fibers that distort the intensity signal measured in reflection, which carries the test mass displacement information.

By using the transfer function given in Equation 2, we convert the displacement measurement from Fig. 6 to acceleration, which is shown in Fig.7.

This plot shows that our monolithic acceleration sensor reaches a measurement noise floor between 100 n$g$/√Hz – 1 µ$g$/√Hz at frequencies above 10 Hz. At observation bandwidths down to 70 mHz, we measure acceleration noise floors of 20 µ$g$/√Hz with this sensor. The current sensitivity limit depends on the fiber noise affecting the test mass displacement measurement.

## 5. Conclusions

We present a fully monolithic and highly compact optomechanical accelerometer that demonstrates a $mQ$-product greater than 50 kg, which is a crucial property for high-sensitivity broad-band acceleration measurements.

The monolithic mechanical oscillator exhibits a quality factor, $Q$, of $4.63\times10^5$. This sensor has a monolithic Fabry-Pérot interferometer with a finesse of 2 that allows either free-space coupling of the light or injection through a GRIN collimator that can be mounted on the sensor itself. Currently displacement noise floors reach levels of $10^{-13}$–$10^{-12}$ m/√Hz above 1 Hz, corresponding to acceleration sensitivities of 100 n$g$/√Hz – 1 µ$g$/√Hz.

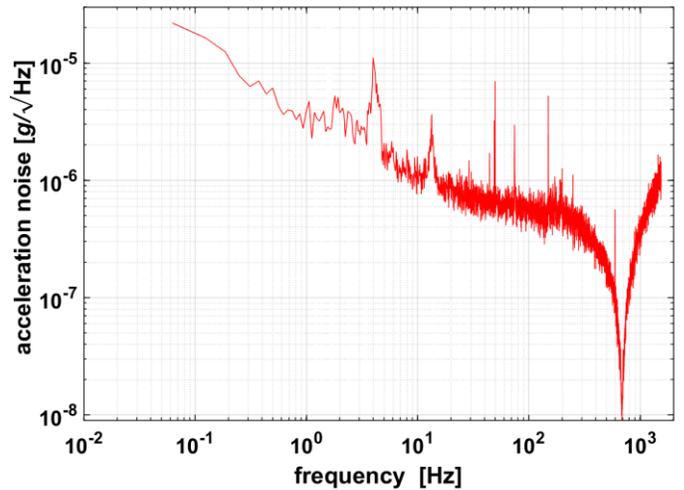

Figure 7. Equivalent acceleration noise floor of the monolithic optomechanical accelerometer.

## 6. Acknowledgements

The authors acknowledge financial support through DFG through CRC 1128 (geo-Q) and its flexible funds project F01.